\documentstyle[epsf,twoside,fleqn,espcrc2]{article}

\newcommand{\AmS}{{\protect\the\textfont2
  A\kern-.1667em\lower.5ex\hbox{M}\kern-.125emS}}

\def\simm#1{\mathop{\vtop{\ialign{##\crcr
        $\hfil\displaystyle{#1}\hfil$\crcr\noalign{\kern0.5pt\nointerlineskip}
        $\sim$\crcr\noalign{\kern0.5pt}}}}\limits}

\def\be{\begin{equation}}
\def\ee{\end{equation}}
\def\bea{\begin{eqnarray}}
\def\eea{\end{eqnarray}}

\title{
\vspace*{-35pt}
{\normalsize \hfill {\sf UTCCP-P-45}} \\
\vspace*{-6pt}
{\normalsize \hfill {\sf Aug.\ 1998}} \\
Full QCD light hadron spectrum from the CP-PACS
\thanks{presented by K.\ Kanaya 
at Lattice98, Boulder, Colorado, USA, 13--18 July 1998.
}
}

\author{CP-PACS Collaboration :\\
 \vspace{1mm}
 S.~Aoki\rlap,\address{Institute of Physics, University of
 Tsukuba, Tsukuba, Ibaraki 305-8571, Japan}
 G.~Boyd\rlap,\address{Center for Computational Physics,
 University of Tsukuba, Tsukuba, Ibaraki 305-8577, Japan}
 R.~Burkhalter\rlap,$^{\rm a,b}$
 S.~Ejiri\rlap,$^{\rm b}$
 M.~Fukugita\rlap,\address{Institute for Cosmic Ray Research,
 University of Tokyo, Tanashi, Tokyo 188-8502, Japan}
 S.~Hashimoto\rlap,\address{High Energy Accelerator Research Organization
 (KEK), Tsukuba, Ibaraki 305-0801, Japan}
 Y.~Iwasaki\rlap,$^{\rm a,b}$
 K.~Kanaya\rlap,$^{\rm a,b}$
 T.~Kaneko\rlap,$^{\rm b}$
 Y.~Kuramashi\rlap,$^{\rm d}$
 K.~Nagai\rlap,$^{\rm b}$
 M.~Okawa\rlap,$^{\rm d}$
 H.P.~Shanahan\rlap,$^{\rm b}$
 A.~Ukawa\rlap,$^{\rm a,b}$ and
 T.~Yoshi\'e$^{\rm a,b}$ }

\begin{document}

\begin{abstract}

We report on an on-going two-flavor full QCD study on CP-PACS using an 
RG-improved gauge action and a tadpole-improved SW quark action. 
Runs are made for three lattice spacings 
$a^{-1}\approx 0.9$, 1.3, and 2.5~GeV on 
$12^3\times24$, $16^3\times32$, and $24^3\times48$ 
lattices. 
Four sea quark masses having $m_{\rm PS}/m_{\rm V} \approx 0.8$--0.6 
are simulated, for each of which hadron masses are evaluated for 
valence quark masses corresponding to 
$m_{\rm PS}/m_{\rm V} \approx 0.8$--0.5. 
Results for hadron and light quark masses are 
presented and compared with those obtained in quenched QCD.

\end{abstract}

\maketitle

\section{Introduction}

Having observed an unambiguous deviation of the quenched light hadron 
spectrum from the experiment\cite{CPPACS-quenched,Yoshie98,ruedi},
we have started, as a logical next step, a systematic full QCD simulation. 
Due to the necessity of working at coarse lattice spacings to cope with 
an increased demand on computer power, we employ an RG-improved 
gauge action combined with a meanfield-improved SW quark action. 
This choice is an outcome of a comparative study of various action combinations
carried out prior to the present work\cite{CPPACS97}. 
In this report we describe the main points of results obtained so far, 
referring to Ref.~\cite{ruedi} for details. 

\section{Parameters of simulation}

\begin{table}[tb]
\setlength{\tabcolsep}{0.2pc}
\caption{Parameters of simulation.}
\label{tab:parameter}
\begin{tabular}{ccccl}
\hline
lattice  & $a^{-1}$[GeV] & 
     $K_{sea}$ & $m_{\rm PS}/m_{\rm V}$ & $N_{\rm conf}$ \\
$\beta/c_{\rm SW}$  & $La$[fm] & & & $\;\;\times N_{\rm sepr}$ \\
\hline
$12^3\times24$ & 0.917(10) & 0.1409 & 0.806(1) & 1250$\times$5 \\
1.80/1.60      & 2.58(3)   & 0.1430 & 0.753(1) & 1000$\times$5 \\
               &           & 0.1445 & 0.696(2) & 1400$\times$5 \\
               &           & 0.1464 & 0.548(4) & 1050$\times$5 \\
\hline
$16^3\times32$ & 1.288(15) & 0.1375 & 0.805(1) & 1400$\times$5 \\
1.95/1.53      & 2.45(3)   & 0.1390 & 0.751(1) & 1400$\times$5 \\
               &           & 0.1400 & 0.688(1) & 1400$\times$5 \\
               &           & 0.1410 & 0.586(3) & 1000$\times$5 \\
\hline
$24^3\times48$ & 2.45(9)   & 0.1351 & 0.800(2) & 250$\times$5 \\
2.20/1.44      & 1.93(7)   & 0.1358 & 0.752(3) & 270$\times$5 \\
               &           & 0.1363 & 0.702(3) & 322$\times$5 \\
               &           & 0.1368 & 0.637(6) & 253$\times$5 \\
\hline
\end{tabular}
\vspace{-0.6cm}
\end{table}

We study QCD with two flavors of sea quarks, identified with 
the degenerate $u$ and $d$ quarks, treating the strange quark 
in the quenched approximation.
Based on test simulations, we choose the 
run parameters listed in Table~\ref{tab:parameter}
corresponding to 
three lattice spacings in the range $a^{-1}\sim 1$--2~GeV.
We employ $12^3\times24$, $16^3\times32$, and $24^3\times48$ lattices 
to keep the spatial size at $L\sim 2.4$~fm.
Simulations are made at sea quark masses corresponding to 
$m_{\rm PS}/m_{V} \approx 0.8$, 0.75, 0.7 and 0.6.

At each sea quark mass, the light hadron spectrum is computed
using valence quark masses corresponding to
$m_{\rm PS}/m_{\rm V} \approx 0.8$, 0.75, 0.7, 0.6 and 0.5.
We smear the quark source with an exponential smearing function 
\cite{CPPACS97}.
All unequal quark mass combinations allowed for degenerate $u$ and $d$ 
sea quarks are taken.
Hadron masses are extracted by uncorrelated fits.
Errors are estimated by the jackknife method  with a bin 
size of 10 configurations generated over 50 HMC trajectories.

\begin{figure}[tb]
\vspace{-2mm}
\begin{center} \leavevmode
\epsfxsize=6.6cm \epsfbox{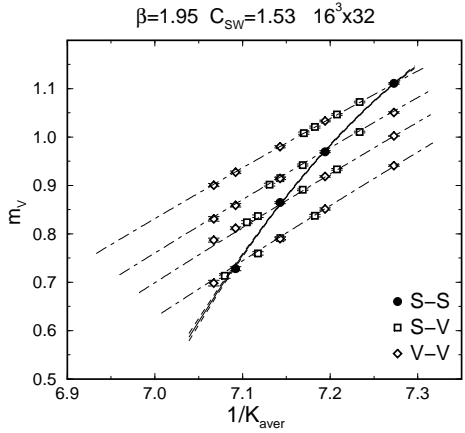}
\end{center}
\vspace{-16mm}
\caption{Vector meson mass at $\beta=1.95$. Lines are results of quadratic 
fit. S and V correspond to $K_{val}=K_{sea}$ and $\ne K_{sea}$,
respectively.}
\label{fig:chiral_vec}
\vspace{-4mm}
\end{figure}

\begin{figure}[tb]
\centerline{\epsfxsize=7.03cm \epsfbox{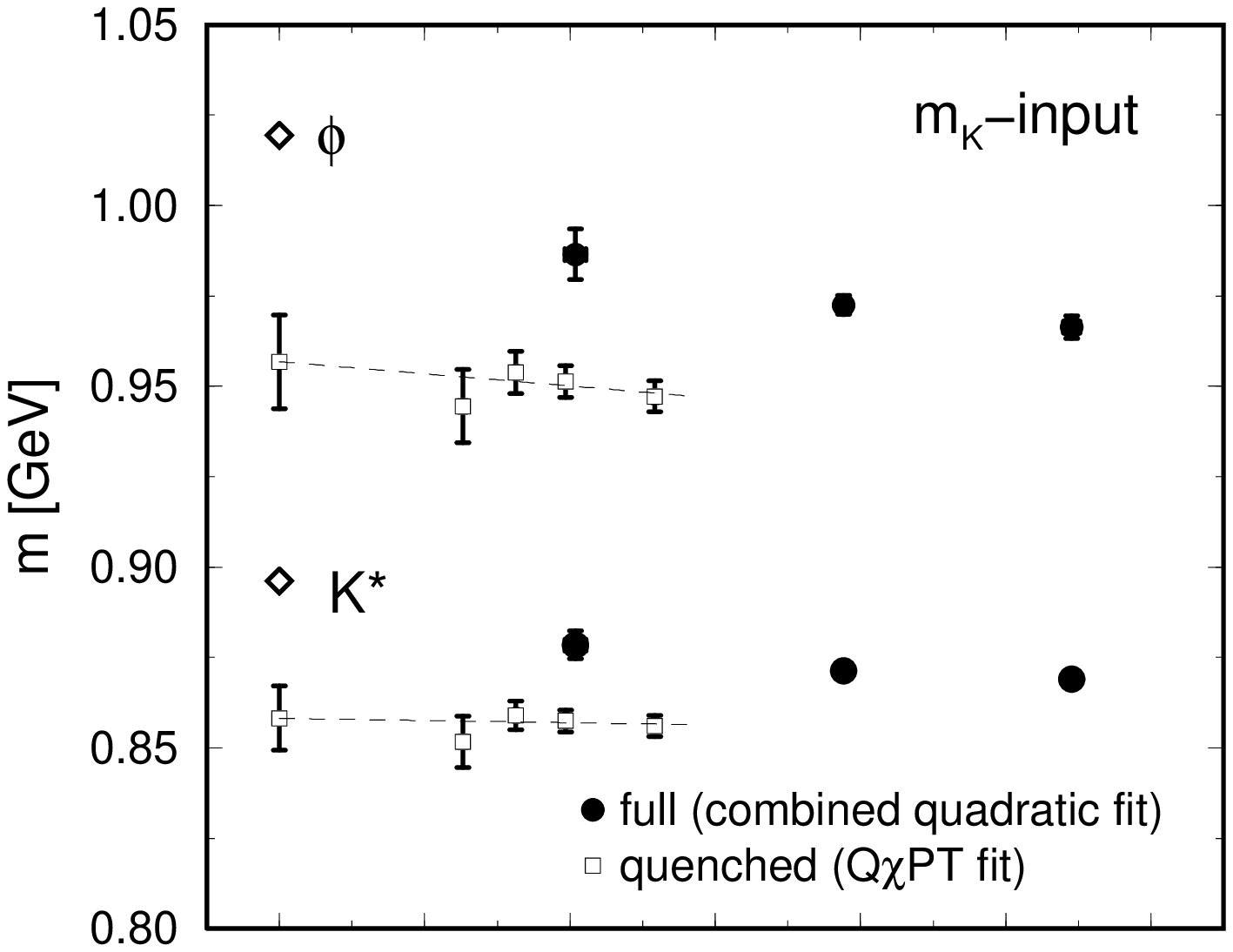}}
\vspace{-4mm}
\centerline{\hspace{0.9mm} \epsfxsize=6.8cm \epsfbox{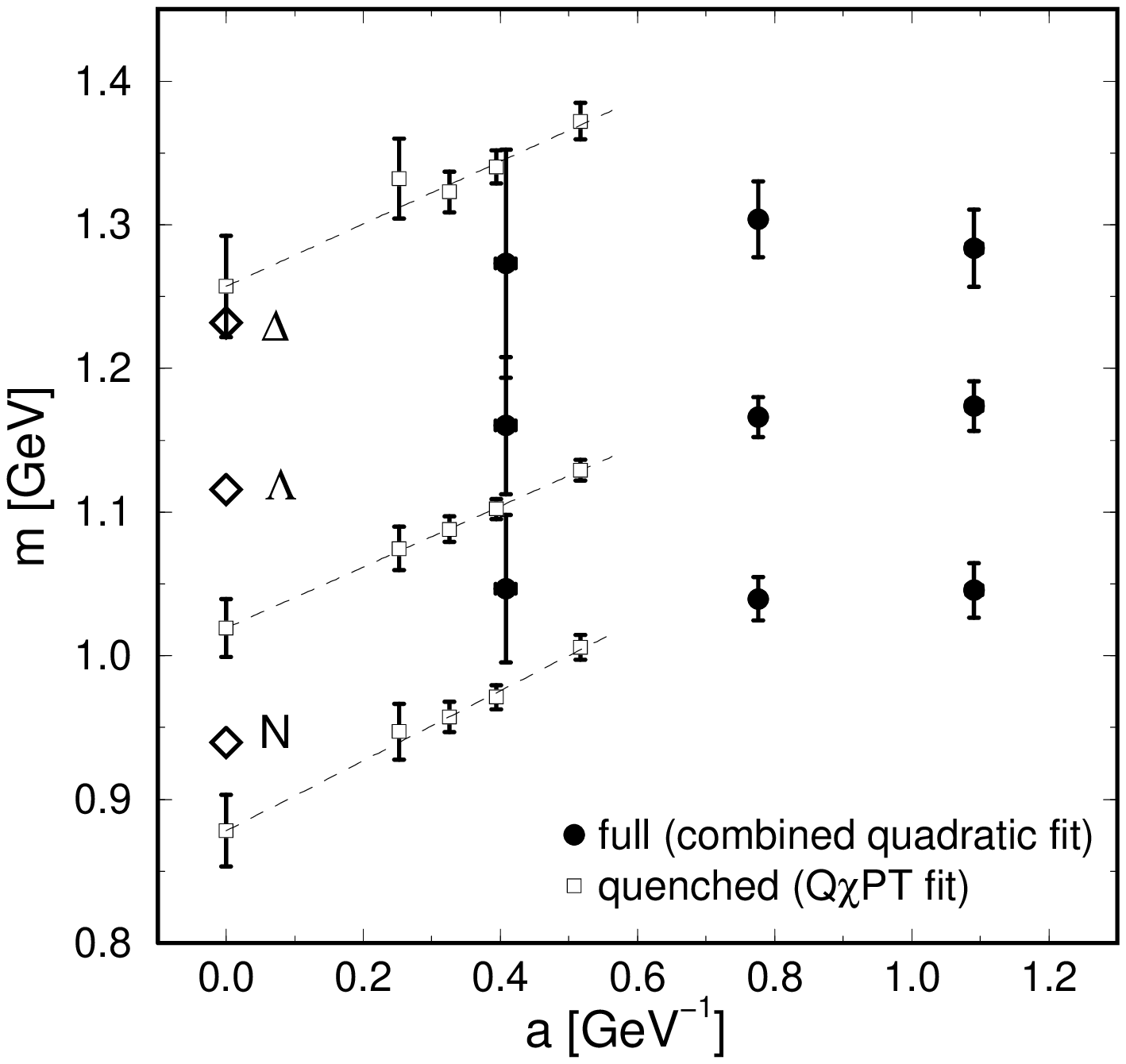}}
\vspace{-12mm}
\caption{Hadron masses as a function of lattice spacing. 
Open squares are the results of quenched QCD with Wilson action 
\protect\cite{Yoshie98}.}
\label{fig:msn_q_cont}
\vspace{-5mm}
\end{figure}

\section{Hadron spectrum}

In Fig.~\ref{fig:chiral_vec} 
we plot the vector meson mass at $\beta=1.95$ as a function 
of the average of $1/K_{val}$ of the valence quark pair of the meson. 
Open points along each dashed lines, which are results of global fits 
described below, are ``partially quenched'' results
calculated for five values of $K_{val}$ for a fixed $K_{sea}$ of sea quark. 
While each of these mass results are almost linear, their slope show a 
variation with $K_{sea}$.  Moreover, the full QCD results satisfying 
$K_{val}=K_{sea}$, which are the four filled points, exhibit a concave
curvature. 
Similar behavior is observed also in other hadrons.

We then make a fit of the mass results 
with a general quadratic ansatz in $1/K_{sea}$ and $1/K_{val}$. 
For pseudoscalar mesons the fit ansatz is
\bea
m_{\rm PS}^2 = B_s \tilde{m}_{sea} + B_v \overline{m}_{val} 
 + C_s \tilde{m}_{sea}^2
 + C_v \overline{m}_{val}^2   \nonumber \\
 + C_{sv} \tilde{m}_{sea} \overline{m}_{val}    
 + C_{12} \tilde{m}_{val(1)} \tilde{m}_{val(2)} \nonumber
\eea
where bare quark masses are defined as
$\tilde{m}_{sea/val(i)} = (K_{sea/val(i)}^{-1} - K_c^{-1})/2$ 
with 
$\overline{m}_{val}\!=\! (\tilde{m}_{val(1)}\!+\!\tilde{m}_{val(2)})/2$
the average valence quark mass.
Similar Ans\"atze without the $C_{12}$ term are used for vector mesons and 
decuplet baryons.  For octet baryons a form inspired from 
chiral perturbation theory
with general quadratic terms of individual $m_{val(i)}$ are employed, 
simultaneously fitting $\Lambda$ and $\Sigma$-like baryons. 

From the fit results, we determine the physical 
light quark point $K_{ud}$ from $m_\pi/m_\rho$,  
and the strange quark point $K_{strange}$ from 
$m_K$, $m_{K^*}$, or $m_\phi$.  The scale is set by $m_\rho$ 
at $K_{ud}$.
Physical hadron masses as a function of the scale are
shown in Fig.~\ref{fig:msn_q_cont} where the results of quenched QCD
with the Wilson quark action\cite{Yoshie98} are also plotted (open symbols).

A very interesting indication in the meson sector is that the present 
two-flavor full QCD result for the hyperfine splitting 
extrapolates to a value 
noticeably closer to experiment than that for quenched QCD
in the continuum limit.
The remaining discrepancy might be due to the quenched treatment of 
the strange quark itself. 

Implications from the baryon results are less clear.  
While scaling violations 
are small, the mass results lie 5--10\% high compared to experiment. 
We note two points in this regard :
(i) the lattice size of 2.4~fm (actually turned out to be 1.93~fm at 
$\beta=2.2$) is smaller compared to 3.0~fm for the quenched case, hence 
finite-size effects can be an issue, 
and (ii) the smallest quark mass 
for full QCD corresponds to $m_{PS}/m_{V}\approx 0.5$ rather than $\approx
0.4$ for quenched QCD, possibly leading to an overestimate of baryon masses.   
Further work matching these points is needed to clarify 
sea quark effects in the baryon mass spectrum.

\begin{figure}[tb]
\centerline{\epsfxsize=6.6cm \epsfbox{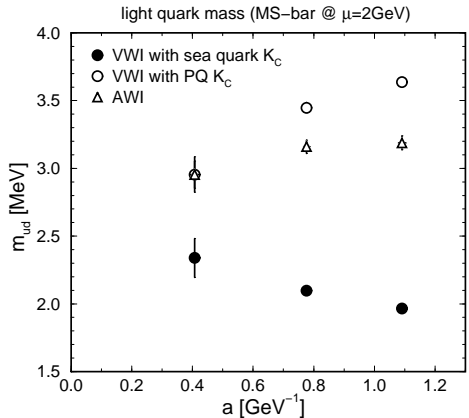}}
\vspace{-13mm}
\caption{Light quark mass 
computed with various definitions.}
\label{fig:light_quark}
\vspace{-5mm}
\end{figure}

\section{Quark masses}

Quark mass can be defined either by vector Ward identity (VWI) 
$m_q^{\rm VWI} \!=\! Z_m m_q$ 
with $m_q \!=\! (1/K_q \!-\! 1/K_{crit})/2$, 
or by axial Ward identity (AWI) 
$Z_A\nabla_\mu\!A_\mu\!=\!2m_q^{\rm AWI} Z_PP$. 
A recent discussion with the VWI definition has been whether to take
the critical value $K_c^{sea}$ of sea quark for $K_{crit}$ or the partially
quenched critical value $K_c^{\rm PQ}$ defined by
$m_{\rm PS}(K_{val})\!=\!0$ with $K_{sea} \!=\! K_{ud}$ 
fixed\cite{SESAMQuark,Gupta}. The former is a natural choice 
for the averaged mass $m_{ud}$ of $u$ and $d$ quarks since sea quarks 
in two-flavor full QCD are identified with them.   
The SESAM Collaboration found, however, that 
this choice leads to a large value of the ratio $m_s/m_{ud}\sim 50$
at $a^{-1}\approx 2$ GeV.  
The value of $m_{ud}$ itself differs by a factor two depending on the two 
choices.  
This effect is negiligibly small for the heavier strange quark. 

In Fig.~\ref{fig:light_quark} we show our results for $m_{ud}$
at 2~GeV in the $\overline{\rm MS}$ scheme.
Renormalization coefficients are taken from one-loop 
tadpole-improved perturbation theory\cite{Taniguchi98}.
The results for the two choices of $K_{crit}$ in VWI, 
while sizably differing at finite lattice spacings, converge to
a common value of 2.5--3 MeV in the continuum limit, and so does the 
AWI quark mass. 
This value is about 40\% smaller 
than 4.6(2) MeV from quenched QCD \cite{Yoshie98}.

The convergence found above also resolves the problem with 
the ratio $m_s/m_{ud}$ for the choice $K_{crit}=K_c^{sea}$; 
while we find $m_{s}/m_{ud}\! \sim\! 50$ at 
$a^{-1}\! \sim\! 0.8$~GeV, the ratio decreases with the lattice spacing,
converging in the continuum limit towards a value
$m_{s}/m_{ud} \approx 25$. 
 
For the strange quark mass itself, 
the main uncertainty arises from the choice of input to fix 
$K_{strange}$.
We find $m_{s}\!\approx\!70$ ($m_K$-input) -- 80 ($m_\phi$-input)
MeV in the continuum limit.
Compared to 115(2) ($m_K$-input) -- 143(6) ($m_\phi$-input) 
obtained in quenched QCD\cite{Yoshie98},
the discrepancy, which might be explained by the quenching of 
strange quark, is smaller, and the values are also smaller.

\vspace{2mm}
This work is supported in part by the Grants-in-Aid
of Ministry of Education
(Nos.\ 08640404, 09304029, 10640246, 10640248, 10740107).
GB, SE, and KN are JSPS Research Fellows. HPS is supported
by JSPS Research for Future Program.


\vspace{-1mm}
\begin{thebibliography}{9}

\bibitem{CPPACS-quenched}
CP-PACS Collab., 
Nucl.\ Phys.\ B (Proc.\ Suppl.) 60A (1998) 14;
{\it ibid.} 63 (1998) 161; T. Yoshi\'e, {\it ibid.} 63 (1998) 3.

\bibitem{Yoshie98}
CP-PACS Collab., presented by 
T.\ Yoshi\'e, these proceedings.

\bibitem{ruedi} R. Burkhalter for CP-PACS Collab., these proceedings.  

\bibitem{CPPACS97}
CP-PACS Collab., 
Nucl.\ Phys.\ B (Proc.\ Suppl.) 60A (1998) 335;
{\it ibid.} 63 (1998) 221.


\bibitem{Taniguchi98}
Y.\ Taniguchi, these proceedings.

\bibitem{SESAMQuark}
SESAM Collab., 
Phys.\ Lett.\ B407 (1997) 290. 

\bibitem{Gupta}
T.\ Bhattacharya and R.\ Gupta, 
Nucl.\ Phys.\ B (Proc.\ Suppl.) 63 (1998) 95.

\end{thebibliography}
\end{document}